## Using Coronal Loops to Reconstruct the Magnetic Field of an Active Region Before and After a Major Flare

A. Malanushenko[1,2], C. J. Schrijver[2], M. L. DeRosa[2], M. S. Wheatland[3]

[1]*Department of Physics, Montana State University, Bozeman, MT, USA*
[2]*Lockheed Martin Advanced Technology Center, Palo Alto, CA, USA*
[3]*Sydney Institute for Astronomy, School of Physics, University of Sydney, Australia*

### ABSTRACT

The shapes of solar coronal loops are sensitive to the presence of electrical currents that are the carriers of the nonpotential energy available for impulsive activity. We use this information in a new method for modeling the coronal magnetic field of AR 11158 as a nonlinear force-free field (NLFFF). The observations used are coronal images around time of major flare activity on 2011/02/15, together with the surface line-of-sight magnetic field measurements. The data are from the Helioseismic and Magnetic Imager and Atmospheric Imaging Assembly (HMI and AIA, respectively) onboard the Solar Dynamics Observatory (SDO). The model fields are constrained to approximate the coronal loop configurations as closely as possible, while also subject to the force-free constraints. The method does not use transverse photospheric magnetic field components as input, and is thereby distinct from methods for modeling NLFFFs based on photospheric vector magnetograms. We validate the method using observations of AR 11158 at a time well before major flaring, and subsequently review the field evolution just prior to and following an X2.2 flare and associated eruption. The models indicate that the energy released during the instability is about $1 \times 10^{32}$ erg, consistent with what is needed to power such a large eruptive flare. Immediately prior to the eruption the model field contains a compact sigmoid bundle of twisted flux that is not present in the post-eruption models, which is consistent with the observations. The core of that model structure is twisted by $\approx 0.9$ full turns about its axis.

### 1. Introduction

Solar energetic events drive space weather, which dictates a need to understand and forecast them. It is important to know *if* a particular active region will erupt, *when* it will erupt, and also how energetic the eruption can be, what the orientation of the magnetic field will be and what the expected velocity of the magnetic cloud formed by the eruption will be.

Modeling of the magnetic field associated with active regions is relevant for answering some of these questions. In the low-$\beta$ corona, the magnetic pressure gradient and the magnetic tension forces dominate over the gas pressure gradient force and the gravity forces (Gary 2001). This means that in an equilibrium, the magnetic field $\mathbf{B}$ needs to be in equilibrium with itself, or in other words, the Lorentz force density

$$\mathbf{F}_L = \frac{c}{4\pi}(\nabla \times \mathbf{B}) \times \mathbf{B} \tag{1}$$

has to vanish everywhere in the corona. This may be rewritten as

$$\nabla \times \mathbf{B} - \alpha \mathbf{B} = 0, \tag{2}$$

where $\alpha$ is a scalar related to how twisted field lines are (e.g., Gold & Hoyle 1960). A field consistent with this equation, along with the divergence-free condition

$$\nabla \cdot \mathbf{B} = 0 \tag{3}$$

is called a nonlinear force free field, hereafter NLFFF (e.g., Nakagawa et al. 1971).

Destabilization of such an equilibrium in the actual low-$\beta$ plasma beyond what the gas forces can balance will result in a reconfiguration of the field with release of magnetic energy. The upper limit on the amount of energy that can be released this way for a particular magnetic field can in principle be derived. Indeed, for given Dirichlet boundary conditions on $\mathbf{B}$, the lowest possible energy is that of a field $\mathbf{P}$ that satisfies

$$\nabla \times \mathbf{P} = 0 \tag{4}$$

(Taylor 1974; Aly 1989). Such a field is called a potential field, which is unique for given boundary conditions. The amount of magnetic energy in excess of the energy of the potential field is therefore called the free energy. The free energy is the upper bound on the amount of energy that could be released, and generally there are additional bounds that could further constraint this value (e.g., helicity conservation Berger 1984).

The free energy is far from being the only quantity relevant for space weather forecasting that can be inferred from NLFFF models. For example, it is believed that the onset of an instability may be determined by how twisted the field is (Hood & Priest 1979) and by how fast the field decays with height (Kliem & Török 2006). Relative orientations of currents in the system may affect what fraction of the free energy can be released in a reconnection (Linton et al. 2001; Linton & Antiochos 2002). The geoeffectiveness of a coronal mass ejection depends on the properties of the ejection, including the orientation of the field relative to the Earth's magnetic field. NLFFF models of particular active regions may provide important inputs for modeling the propagation of coronal mass ejections through the heliosphere.

NLFFF modeling can provide estimates of these important quantities, but constructing such a model for a given region on the Sun is an extremely challenging task. One of the difficulties is the availability of the boundary conditions for the NLFFF problem. To solve Equations (2) and (3) in a given domain $\mathcal{V}$, boundary conditions must be set on $\mathbf{B} \cdot \hat{\mathbf{n}}|_{\partial\mathcal{V}}$, and the value of $\alpha$ must be set on each field line[1] (Bineau 1972). At the lower boundary, $z = 0$, the latter can be derived if the full vector $\mathbf{B}$ is known:

$$\alpha = J_z/B_z = \frac{1}{B_z}\left(\frac{\partial B_y}{\partial x} - \frac{\partial B_x}{\partial y}\right). \tag{5}$$

One of the problems immediately arising is that vector magnetograms (e.g., Scherrer et al. 2012) are measured at the photospheric level, where the low-$\beta$ approximation is not correct and therefore Equation (2) does not apply. In other words, $\alpha$ *does* change along field lines between the photosphere, where it is measured, and the lower boundary of the low-$\beta$ corona, which Equation (2) is meant to model (see Démoulin et al. 1997, for an extensive discussion of this issue).

An additional problem is that $\alpha$ derived using Equation (5) is affected by measurement uncertainties in the horizontal components of $\mathbf{B}$. This has more impact than it might seem at first sight: field lines in a NLFFF must connect points with the same $\alpha$ on positive and negative polarities at the lower boundary so the boundaries must have equal amounts of incoming and outgoing magnetic flux for each value of $\alpha$. Noise in $\alpha$ at the lower boundary, the fact that the photosphere is not force-free, and limits to the field of view prevent this condition from being generally satisfied (Aly 1989).

One approach to deal with these problems is to modify the boundary data to make them consistent with the force-free conditions. These approaches include the pre-processing of the data (e.g., Wiegelmann et al. 2006, 2008), or, alternatively, formulating a well-posed problem given the uncertainties in the measurements (Amari & Aly 2010; Wheatland & Régnier 2009; Wheatland & Leka 2011). Various methods for solving the NLFFF problem subject to the prescribed boundary data are available, for example, vertical integration (Nakagawa 1974; Wu et al. 1990), magnetofrictional relaxation (e.g., Mikic & McClymont 1994; van Ballegooijen 2004; Valori et al. 2005), optimization (e.g., Wiegelmann 2004) and the Grad-Rubin method (e.g., Sakurai 1981; Amari et al. 1997; Amari et al. 1999; Wheatland 2007). However, when applied to the same photospheric data, different methods have been shown to yield results often inconsistent with each other and with coronal features (DeRosa et al. 2009). Moreover, the same method may produce two different answers, depending on whether the information about $\alpha$ is drawn from the positive or from the negative polarity at the photosphere, even if the data are pre-processed. There are methods to deal with this issue (e.g., Wheatland & Régnier 2009), but that comes at the price of the model field being not fully consistent with the observed vector field at the lower boundary (pre-processing *per se* also implies that the data are altered).

In this manuscript, we use a different approach: we make use of another source of information about $\alpha$, bypassing the problems associated with measuring $\alpha$ at the lower boundary altogether. The values of $\alpha$ along some field lines in the corona may be evaluated using coronal loop observations in the Extreme Ultraviolet (EUV), as was demonstrated by Malanushenko et al. (2009b). In our previous study (Malanushenko et al. 2012) we developed a method to construct a NLFFF based on these $\alpha$ data, along with the vertical component of $\mathbf{B}$ at the lower boundary. We tested the method (called Quasi Grad-Rubin, or hereafter QGR) on synthetic data and on known fields and found that it produces generally reliable results. In particular, it is capable of reproducing large-scale features (the structure of currents and the shape of field lines) and can be used to evaluate the free energy of the field. In this manuscript, we take the next logical step and apply the QGR algorithm to real data. We briefly describe the algorithm in Section 2, then we validate it in Section 3 by creating several field models in a quiescent region within short time intervals. In Section 4 we apply the QGR method to create models of the field of the same active region in a time interval when this active region exhibits major activity. Finally, in Section 5 we summarize the results of both validation of the QGR method on the real data, and on the results of the study of major eruptive activity with its use.

[1]By taking the divergence of both sides of Equation (2) it follows that $\mathbf{B}\cdot\nabla\alpha = 0$, which implies that $\alpha$ has to be constant on each field line.

AIA 171Å image on 2011-02-15 04:00 | Loops identified in 6 AIA channels | Best-fitting lines of constant-$\alpha$ fields

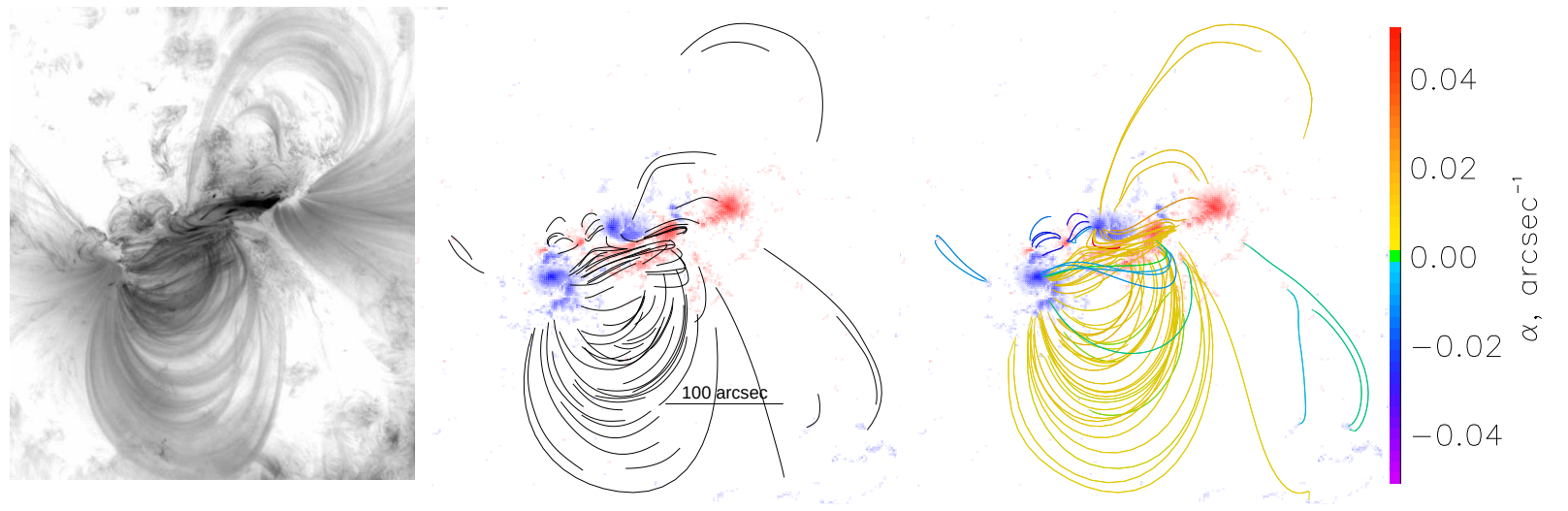

Fig. 1.— Coronal loops as constraints for magnetic modeling. Loops observed in EUV channels (the left panel shows an AIA 171Å image) are approximated by smooth curves (the middle panel compiles loops visible in several different AIA channels) which are fit by field lines of linear force-free fields (as shown in the right panel, where different colors correspond to different values of $\alpha$).

## 2. The Quasi-Grad-Rubin (QGR) Method

In the Grad-Rubin (hereafter GR) iteration, Equation 2 is solved in a domain $\mathcal{V}$ subject to boundary condition on the normal component of $\mathbf{B}$ on the boundary, $\mathbf{B} \cdot \hat{\mathbf{n}}|_{\partial\mathcal{V}}$, and the values of $\alpha$ on some surface $S$ which all field lines pass through. The iteration begins with an initial guess for $\mathbf{B}$, for example, $\mathbf{B}^{(0)} = \mathbf{P}$. Then, $\alpha$ is propagated from the surface $S$ (where its value is known) to every point in $\mathcal{V}$ along field lines of $\mathbf{B}^{(0)}$, thus creating a volume-filling $\alpha^{(0)}$. In the next iteration, $\mathbf{B}^{(1)}$ is obtained by solving the equation $\nabla \times \mathbf{B}^{(1)} = \alpha^{(0)}\mathbf{B}^{(0)}$ subject to the prescribed boundary conditions for $\mathbf{B}$. The values of $\alpha$ are then propagated from the surface along field lines of $\mathbf{B}^{(1)}$ yielding a volume-filling $\alpha^{(1)}$, and so on. The iteration repeats until $\mathbf{B}^{(n)} \approx \mathbf{B}^{(n-1)}$ and therefore Equation (2) is satisfied to within a desired tolerance. This procedure was theoretically proven to converge to a force-free solution for small $\alpha$ and for well-posed boundary conditions (Bineau 1972).

The QGR iteration involves modification of a particular implementation of the GR iteration called cfit (Wheatland 2007). In the cfit iteration scheme, $\mathbf{B}^{(i+1)}$ is found in terms of the vector potential, $\mathbf{B}^{(i+1)} = \nabla \times \mathbf{A}^{(i+1)}$, assuming the Coulomb gauge, by solving the Poisson equation $\nabla^2\mathbf{A}^{(i+1)} = -\alpha^{(i)}\mathbf{B}^{(i)}$. A 2-D Fourier solution to the Poisson problem is used. The fact that $\mathbf{B}^{(i+1)}$ is *calculated* as a curl of some other vector ensures $\nabla \cdot \mathbf{B}^{(i+1)} = 0$ to within numerical error. The volume-filling $\alpha^{(i)}$ is found at each point of the domain by tracing a field line in $\mathbf{B}^{(i)}$ from this point in both directions until the field line reaches the lower boundary at the positive (or negative) polarity (the sign of the polarity is selected prior to initializing the iterations). If the field line leaves the domain through a side or the top boundary, $\alpha^{(i)}$ at this point in the domain is set to 0. In cfit, the field at the side and the top boundary changes through the iterations, however, no electric currents are allowed to cross these boundaries. The field is split into a potential and a non-potential parts, $\mathbf{B}^{(i+1)} = \mathbf{P} + \mathbf{B}_c^{(i+1)}$, where only the second term is influenced by currents. The potential field $\mathbf{P}$ remains constant at all points, including on the boundary, and the non-potential part $\mathbf{B}_c^{(i+1)}$ can be obtained without explicitly setting boundary conditions, since the right hand in $\nabla^2\mathbf{A}^{(i+1)} = -\alpha^{(i)}\mathbf{B}^{(i)}$ is known at all points in the volume at $i$-th iteration. This is explained in more detail in Wheatland (2007).

The QGR is algorithmically very similar to cfit, however, it was modified to deal with a different kind of input data: $\alpha$ is no longer set as a *boundary condition* on a surface. Instead, $\alpha$ is constrained to maintain specified values at a set of specific points in the volume. The term "boundary conditions" is obviously inapplicable for such constraints, so we hereafter refer to these as to "constraints in the volume". These constraints in the volume are obtained from EUV observations of coronal loops, by finding, for each coronal loop, a field line in a constant-$\alpha$ field (that is, a force-free field with $\alpha$ =const at all points, also referred to as linear force-free field, hereafter, LFFF) which best matches the observed loop in the plane of sky (Malanushenko et al. 2009b). The $\alpha$ value of this LFFF is expected to match approximately the actual value of $\alpha$ in the coronal NLFFF, and the field line found this way is in turn expected to approximately match the 3D trajectory of the actual coronal loop, for reasons demonstrated in Malanushenko et al. (2009b, 2011). Figure 1 shows an example of coronal loops observed in the EUV and the constraints in the volume obtained from these loops. These data *per se* are not a NLFFF model: they are merely a sparse set of trajectories approximating the observed loops and the derived values of $\alpha$ on these trajectories. An additional procedure is needed to construct a volume-filling force-free field from these data.

The main difference between QGR and cfit is in the step where the volume-filling $\alpha^{(i)}$ is calculated. The way this step is performed in cfit is explained above. QGR maintains the same general idea: $\alpha^{(i)}$ should be propagated along field lines of $\mathbf{B}^{(i)}$. However, instead of using values of $\alpha$ on a boundary surface, in QGR, $\alpha$ is *averaged* along the volume-filling values from the previous iteration, $\alpha^{(i-1)}$. The initial guess, $\alpha^{(-1)}$, is a piecewise-constant cube where the value of $\alpha$ at each point is determined from the value of $\alpha$ at the nearest constraint in the volume. $\alpha^{(0)}$ is obtained by averaging the initial guess along field lines of $\mathbf{B}^{(0)} = \mathbf{P}$. This is implemented in a manner similar to that in cfit, as follows. A field line is traced from each point. If the field line leaves the domain through a side or the top boundaries, the value of $\alpha^{(i)}$ at this point is set to 0. Otherwise, the value of $\alpha^{(i)}$ at this point is set to the average of $\alpha^{(i-1)}$ along this field line traced in $\mathbf{B}^{(i)}$. Before this tracing step, the values of $\alpha$ along the loop trajectories are re-set to the constraining values determined by the EUV observations. Note that on each iteration, the step where $\alpha$ is averaged along the field lines is taken prior to the step where the field is uncurled. This ensures that in $\nabla^2\mathbf{A}^{(i)} = -\alpha^{(i-1)}\mathbf{B}^{(i-1)}$, the right part of the equation is divergence free.

In this work, we introduce a minor modification to the scheme from Malanushenko et al. (2012). While tracing field lines from a given point, we set an upper length limit for which a field line is integrated, and $\alpha$ is only averaged over this segment of a complete field line. For the models in this paper, the length limit is set to 100 pixels, or about 40% of the horizontal extent of the domain. This is done to make the solution more stable during the initial steps, where large variations of $\alpha$ along field lines are anticipated. This is expected to make little difference after many iterations (typically a few hundred).

There is one further important detail. The iterations in QGR are damped in order to stabilize the solution and account for possible uncertainties in the input data. This is done by introducing a damping parameter, $\alpha_{thr}$. In the step where the new volume-filling cube $\alpha^{(i)}$ is calculated, the value at a given point is not changed from $\alpha^{(i-1)}$ (or from a constraint in the volume, if this is where one of these constraints is set), if the magnitude of the change is smaller than the value of $\alpha_{thr}$. The latter is set, by default, to the estimated numerical error expected in a given problem (as explained in Malanushenko et al. 2012). The value of $\alpha_{thr}$ may be gradually increased where uncertainties in the input data make the constraints inconsistent with a force-free solution. This enables the field to be kept as force-free as possible, while matching the constraints, again, as closely as possible.

It is recognized that the LFFF fitting method used by Malanushenko et al. (2009b) underestimates values of $\alpha$ by some constant factor, which is found to be slightly different for different model fields, when tested on synthetic data. Malanushenko et al. (2012) showed that the QGR modeling yields a better match to the loop data when this underestimation is accounted for, that is, when the QGR model is constructed with the values $\alpha' = f\alpha$, where $f$ is the constant underestimation factor for the LFFF fitting method. Moreover, Malanushenko et al. (2012) showed that in a sequence of QGR models which have the same set of constraints in the volume, but in each model the values of the constraints in the volume are set to $\alpha' = f'\alpha$ with $f'$ =const in a given model, the model which best matches the observed loops is the one with $f' = f$. This fact allows us to evaluate $f$ if it is not known ahead of time, by constructing a sequence of QGR models for different values of $f$, and determining, which value yields the best match to the loops. As was shown in Malanushenko et al. (2012), this QGR model will also be the closest match to the field simulated.

We define $\overline{d}$, a measure of the quality of fit of the QGR model to the loop trajectories. This is derived as the average distance between the loops and the QGR model field lines projected onto the plane of sky (see Malanushenko et al. 2009b). Here, for each $i$-th loop, we use an automated algorithm that looks for a best-fitting QGR model field line in the vicinity of the $i$-th LFFF loop track. For each of the multiple field lines initiated at a thin volume surrounding the track (that is, from all points closer than some distance, which we set to 10 pixels), the one that is the closest match (in terms of the distance to the loop) is kept, and the distance $d_i$ is stored. In the previous study, we considered the average $d_i$ as the metric for how well a given field model matches the loops. However, we note that within the distribution of $d_i$, several of the values are typically much larger than the rest, due to poor fits, and these large values unduly influence the mean. In the current study, we use the *median* of the distribution, making $\overline{d}$ less sensitive to the values of a few outlying data points. For the given QGR model constructed with a particular $f$ value, $\overline{d}$ is the median of all $d_i$, and for a set of such models, $\overline{d}(f)$ is minimized, yielding $\overline{d}_{best}$ and $f_{best}$. The QGR model corresponding to $f_{best}$ is considered to be the final product for a given dataset. We quantify how well this model matches the loops by the value $\overline{d}_{best}$. For context, we also report $\overline{d}_{pot}$, which is the same metric for the potential field. To evaluate $\overline{d}$, we use a slightly larger set of the loop tracks than is used for QGR constraints in the volume. As explained in Malanushenko et al. (2011), every LFFF loop fit is visually examined and assigned a subjective rating of A (excellent fit), B (acceptable fit), C (questionable fit) or D (failure). Only A and B quality loops are used as constraints in the volume. However, for the procedure of determining $\overline{d}$ we also include the C-quality loop tracks as well, as only their approximate location is used to initiate the field line search.

### 2.1. Measures of how Force-Free a Field Is

Two metrics for the quality of a numerical approximation to a NLFFF are how force-free the field is (that is, how well Equation (2) is fulfilled in the computational domain), and how divergence-free it is (e.g., Schrijver et al. 2006). As we explain above, the QGR method solves for the vector potential $\mathbf{A}$. The model field $\mathbf{B}$ is calculated by taking the curl of the vector potential: $\mathbf{B} = \nabla \times \mathbf{A}$, therefore, within the same finite difference scheme, $\nabla \cdot \mathbf{B} = 0$ is automatically fulfilled subject to numerical error. Therefore, we only examine how well the force-free condition is met.

One metric that has been used is the current-weighted sine of the angle between $\mathbf{B}$ and the

current density **J**:

$$\text{CWsin} = \frac{\sum |\sin\mu||\mathbf{J}|}{\sum |\mathbf{J}|}, \tag{6}$$

where $\mu$ is the angle between $\mathbf{B}$ and $\mathbf{J} = (c/4\pi)\nabla\times\mathbf{B}$, that is, $|\sin\mu| = |\mathbf{B}\times(\nabla\times\mathbf{B})|/(|\mathbf{B}||\nabla\times\mathbf{B}|)$ (e.g., Metcalf et al. 2008).

While CWsin $\ll 1$ implies the field is force-free, an increased value of CWsin does not necessarily indicate the opposite. It may be large if the domain is comprised of two subdomains, one of which is perfectly free of currents ($\mathcal{V}_\mathrm{P}$), and the second one is perfectly force-free ($\mathcal{V}_\mathrm{ff}$), as the case is often in QGR and other methods in which open flux (flux that leaves the volume through the side and top boundaries) is required to be current-free. Indeed, in this case $|\sin\mu|$ is small in the force-free domain which has non-zero currents. In the current-free domain $\nabla\times\mathbf{B}$ is non-zero due to numerical error and the magnitude of this vector is small, but the current due to errors has random orientations and so $|\sin\mu|$ is non-zero (in fact, the distribution of $|\sin\mu|$ for the potential field peaks at 1, which implies $\mu = \pi/2$ on average). It is normally accepted that the normalization by $|\mathbf{J}|$ (as in Equation (6)) accounts for the current-free regions. However, when a significant portion of the volume is current-free, CWsin may become dominated or seriously influenced by such noise values and so it may not be an informative measure.

We propose an alternative metric for force-freeness, which is insensitive to the presence of a current-free sub-volume. Rather than using $\sin\mu$ as the metric and weighting it with current density, we propose to use the Lorentz force itself, and normalize it by the magnitude of its physically meaningful components. The Lorentz force may be written as (Malanushenko 2010):

$$\mathbf{F}_\mathrm{L} = -\frac{c}{8\pi}\nabla_\perp B^2 + \frac{c}{4\pi}B^2\frac{d\hat{\mathbf{b}}}{db} \equiv \mathbf{F}_\mathrm{mp} + \mathbf{F}_\mathrm{mt}, \tag{7}$$

where $\mathbf{B} = B\hat{\mathbf{b}}$, $b$ is the distance along a field line and $\nabla_\perp$ is the component of the gradient perpendicular to $\hat{\mathbf{b}}$. The first term is commonly referred to as the magnetic pressure gradient force and the second term is the magnetic tension force, by mathematical analogy with elastic tension. Therefore, the force balance between magnetic, gravity and pressure forces may be written as

$$\mathbf{F}_\mathrm{mp} + \mathbf{F}_\mathrm{mt} + \mathbf{F}_\mathrm{g} + \mathbf{F}_\mathrm{p} = 0, \tag{8}$$

where $\mathbf{F}_\mathrm{g}$ and $\mathbf{F}_\mathrm{p}$ are the gravity and pressure gradient forces respectively.

Estimates for coronal parameters suggest that $|\mathbf{F}_\mathrm{mp}| \gg |\mathbf{F}_\mathrm{p}|$ (Gary 2001), and that $|\mathbf{F}_\mathrm{g}| \approx |\mathbf{F}_\mathrm{p}| \ll |\mathbf{F}_\mathrm{mp}|$ is reasonable (Schrijver & Zwaan 2000). Hence, Equation (8) implies that $\mathbf{F}_\mathrm{mp} + \mathbf{F}_\mathrm{mt} \approx 0$, which is another way to write the force-free Equation (2). The accuracy of the force-free approximation is determined by the pressure gradient and gravity forces, $\mathbf{F}_\mathrm{p} + \mathbf{F}_\mathrm{g}$, which could balance a small, but not vanishing Lorentz force.

This suggests that a suitable measure of whether $\mathbf{F}_\mathrm{L}$ is small is comparison with $\mathbf{F}_\mathrm{p} + \mathbf{F}_\mathrm{g}$. However, in NLFFF models, $\mathbf{F}_\mathrm{g}$ and $\mathbf{F}_\mathrm{p}$ are absent and it might appear at first sight that such comparison can not be made. This is not entirely true: by writing Equation (2) a certain assumption about the magnitude of these forces is already made, as we explain above.

To summarize, if the the Lorentz force does not vanish completely, it is still possible for the low-$\beta$ plasma to be in equilibrium, but an upper boundary on the magnitude of the Lorentz force can be set:

$$|\mathbf{F}_{\rm L}| \approx |\mathbf{F}_{\rm p} + \mathbf{F}_{\rm g}| \leq |\mathbf{F}_{\rm p}| + |\mathbf{F}_{\rm g}| \approx |\mathbf{F}_{\rm p}| + |\mathbf{F}_{\rm p}| \ll |\mathbf{F}_{\rm mp}| + |\mathbf{F}_{\rm mp}| \approx |\mathbf{F}_{\rm mp}| + |\mathbf{F}_{\rm mt}|. \qquad (9)$$

Note that we do *not* claim this has to be true everywhere in the corona. Rather, we demonstrate that this is a direct consequence of the basic underlying assumption which is made when NLFFF modeling is used. That is to say, Equation (9) might not hold, but in these circumstances the NLFFF modeling should not be used at all. Also note that Equation (9) is a necessary but not sufficient condition to enable a low-$\beta$ equilibrium.

Hence we propose to introduce the new metric for how force-free a field is:

$$\xi = \frac{1}{N}\sum_{i=1}^{N} \frac{|\mathbf{F}_{\rm L}|}{|\mathbf{F}_{\rm mp}| + |\mathbf{F}_{\rm mt}|}. \qquad (10)$$

This is the average of the Lorentz force relative to its components at each point of the domain over the total $N$ points of the domain. Just like CWsin, $\xi$ is a dimensionless number, and $0 \leq \xi \leq 1$. Unlike CWsin, the quantity $\xi$ is not sensitive to the absence of currents. The condition $\xi \ll 1$ is a necessary condition for use of force-free modeling, and $\xi \sim 1$ implies substantial Lorentz forces in the volume.

Further in the text we provide both $\xi$ and CWsin metrics for the field models, but CWsin is given mostly for consistency with the existing literature. We believe that a large (and irregularly shaped) current-free portion of the volume might impact how informative CWsin is.

## 3. Validation of the Method

The basic method was tested on known model fields by Malanushenko et al. (2012). Here, we perform an additional test by investigating the sensitivity of the QGR method to uncertainties in the input data. For example, we need to verify that when different sets of loops are drawn from the same field, they lead to similar QGR solutions. Malanushenko et al. (2012) showed on synthetic data, that when different sets are comprised of coronal loops selected at random, the solution does not depend strongly on the particular set used. However, the selection mechanism which "highlights" coronal loops on the Sun might not be completely random, and we test here how this may influence the modeled field.

The experiment we use to validate the QGR method is as follows. We select an active region for which we don't observe strong changes in the coronal field over a certain interval longer than a typical lifetime of a loop. Within this interval, we construct several field models. This is designed to mimic the process of reconstructing the same (or very similar) states of the corona from several different sets of coronal loops.

We chose AR 11158 for this and the next experiments, because this is a well-studied active region[2], so comparisons with other studies is possible. We select a time interval of one hour during which there was no significant activity in the EUV, and confirm in AIA observations that the coronal field changed relatively slowly. The light curve and the magnitude of the running difference in the left column of Figure 2 confirm the lack of major EUV changes. Within this time interval, we construct four QGR models of the field corresponding to four time frames at 20 minute cadence. These times are marked in Figure 2 with arrows. The cadence was chosen to be of the order of a typical lifetime of a loop (20 minutes, according to Longcope et al. 2005). This experiment imitates reconstruction of approximately the same field based on several different (although overlapping) sets of observed loops.

The coronal loops (used to derive the constraints in the volume) are obtained from full-resolution 94Å, 131Å, 171Å, 193Å, 211Å and 335Å AIA channels, yielding $\approx 130$ manually traced loops at each time. These loops were fit by LFFF field lines, as explained in Section 2, with $\approx 3/4$ success rate (each fit was visually examined and obviously unacceptable fits were discarded). The HMI line of sight magnetograms used as a lower boundary are downsampled to about 1.8 arcsec per pixel to reduce computational time. The computational domain is the same for all models of this active region. It is chosen to be large enough to encompass the AIA loops at the studied times, namely, $460 \times 460 \times 229$ arcsec or $256 \times 256 \times 128$ pixels. We use the common value of $\alpha_{thr} \approx 2 \times 10^{-6}$arcsec$^{-1}$ for the models. For the first model in the sequence, we compute several realizations with different values of $f$ within the range $f = 1 - 5$ with steps of $\Delta f = 0.5$. The optimal value of $f$ which results in the best fit to the coronal loops is found to be $f \approx 1.5$, using the procedure explained in Section 2. For each of the subsequent three models, values of $f$ in the range $f = 1 - 3$ are used. The iterations are typically terminated when the average relative change in the field between consecutive iterations drops to the order of $10^{-6}$.

The resulting four QGR models are very similar to one another, even though they are based on different loop traces. The top row in Figure 3 compares the traced coronal loops to the best-fitting field lines in the corresponding QGR model. All four models have very similar features as shown in the on-line only Figure S1. The first four rows in Table 1 summarize properties of each of the four models. The average free energy of the models is $E_F \approx (5.5 \pm 1) \times 10^{30}$ erg. The value $\overline{d}_{best}$ quantifies how well the field lines of the QGR model match the observed loops (the procedure for calculating $\overline{d}$ is described in Section 2). For context, we also show the value $\overline{d}_P$, which quantifies how well potential field lines match the observed loops; $\overline{d}_{best} < \overline{d}_P$ for all four models, which means that the QGR models match the observed coronal features better than the potential field lines. We also show the structure of the electric currents for the first model in the sequence (Figure 4, top row). We separately integrate the magnitude of currents for $\alpha > 0$ (left image) and $\alpha < 0$ (right image) domains. The same plots for all four models are also very similar, as shown in the on-line only Figure S2.

[2]In fact, it might be *the* most well-studied active region in the SDO era. At the time this manuscript was written, the NASA's Astrophysics Data System Bibliographic Services listed over 30 peer-reviewed articles that mention AR 11158 in the abstract. An excellent review of the major findings by many of these studies is given in Section 3 of Tziotziou et al. (2013).

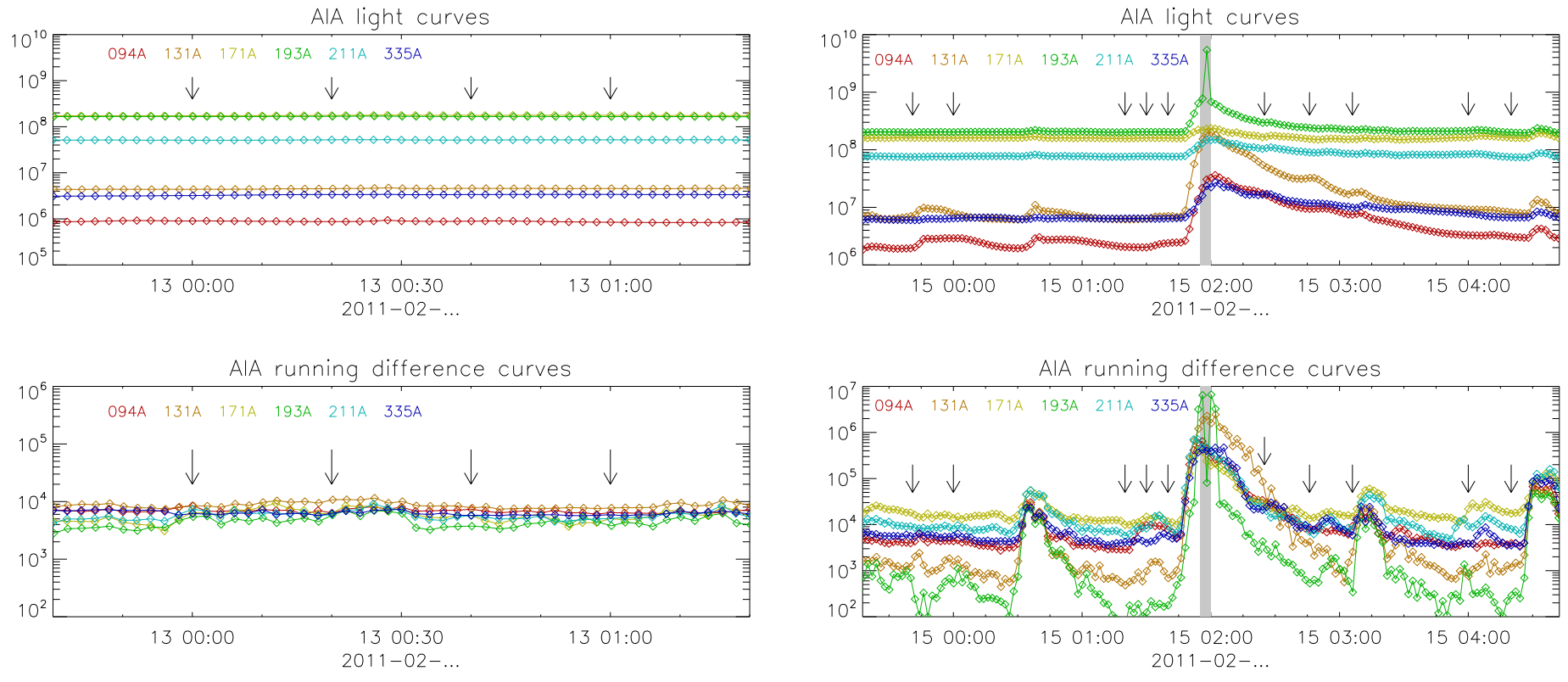

Fig. 2.— Top panels: light curves (in arbitrary units) for AR 11158 in several AIA channels, over the region used to trace coronal loops. The arrows denote the times for which the model fields are created. Bottom panels: "light curves" of the magnitude of running difference images. This figure shows that on 2011-02-13 at 00:00 to 01:00 there was little activity in the AR 11158, and that several major bursts of activity took place later on, on 2011-02-15 between 00:00 and 04:00, including an X2.2 flare (the approximate peak time of this flare is here and on further plots marked with the light gray band). The running difference curve was designed to capture large-scale changes even if they are not associated with excessive brightening (such as filament eruptions). It is calculated as follows. At each pixel of the image, we calculate the standard deviation of the running difference (assuming the mean is zero) at this pixel. This allows to evaluate "typical" and "atypical" changes at a given location. The running difference images are then normalized to this image of standard deviations. For the running difference light curve, we integrate the magnitude of this normalized running difference over the areas where it is larger than $3\sigma$.

We also quantitatively compare the model fields to each other using the following metrics:

- $\mathrm{C_{CS}} = \frac{1}{N}\sum\left[\mathbf{B}_1 \cdot \mathbf{B}_2/(|\mathbf{B}_1||\mathbf{B}_2|)\right]$ (where $N$ is the number of points in the domain), which is the average cosine of the angle between the fields $\mathbf{B}_1$ and $\mathbf{B}_2$;

- $C_{vec} = (\sum \mathbf{B}_1 \cdot \mathbf{B}_2)/((\sum |\mathbf{B}_1|^2)(\sum |\mathbf{B}_2|^2))^{1/2}$, which is the same as $C_{CS}$ but with increased weight in regions of stronger field;

- $1 - E_{m1} = 1 - \frac{1}{N} \sum [|\mathbf{B}_1 - \mathbf{B}_2| / (|\mathbf{B}_1| + |\mathbf{B}_2|)]$, which is the average relative difference between $\mathbf{B}_1$ and $\mathbf{B}_2$;

- $1 - E_{n1} = 1 - (\sum |\mathbf{B}_1 - \mathbf{B}_2|) / (\sum [|\mathbf{B}_1| + |\mathbf{B}_2|])$, which is the same as $1 - E_{m1}$ but with increased weight in regions of stronger field.

The first two metrics are used in the existing literature (e.g., Metcalf et al. 2008). The second two are slight modifications of the metrics $1 - E_n$ and $1 - E_m$, which have been used before, but here they have been normalized to a value between 0 and 1, and have been modified to treat the fields $\mathbf{B}_1$ and $\mathbf{B}_2$ symmetrically, i.e., $E_{n1}(\mathbf{B}_1, \mathbf{B}_2) = E_{n1}(\mathbf{B}_2, \mathbf{B}_1)$ and similarly for $E_{m1}$.

For a given metric, a higher value means that the two compared fields are more similar. We calculate these metrics for pairs of consecutive QGR models. For reference, we compute the metrics for pairs of consecutive potential fields. The third similarity comparison is that between a QGR model field and a corresponding potential field. The results of the quantitative comparison are shown in Figure 5. We find that: (1) comparing a set of successive potential fields with each other yields metrics of order 0.95 or higher; (2) the metrics resulting from the comparison amongst the set of QGR models are almost as high as those from the potential field set; (3) when comparing the QGR models to the potential field models at each time, the metrics are all lower than the comparisons between potential or QGR models to themselves. These findings demonstrate that the QGR models are more similar to each other than to the corresponding potential fields, indicating a degree of consistency within the set of QGR models.

Table 1: Parameters of the QGR field models in the quiescent phase of AR 11158. Notes: [a] – potential f
energy: energy of the QGR model field less the potential field energy. [c] – total energy of the QGR model, 
as defined by Equation (6). [e] – as defined by Equation (10). [f] – the value of $f$ which minimizes $\overline{d}(f)$, as ex
[g] – the minimal value $\overline{d}(f)$. [h] – the value of $\overline{d}$ for the potential field. [i] – the number of loops traced manua
EUV channels. [j] – the fraction of the total loops which were successfully fit by the constant-$\alpha$ field lines.

| 2011-02-... | $E_P$ [a] | $E_F$ [b] | $E_{tot}$ [c] | $E_{tot}/E_P$ | CWsin [d] | $\xi$ [e] | $f_{best}$ [f] | $\overline{d}_{best}$ [g] | $\overline{d}_{pot}$ [h] | |
|---|---|---|---|---|---|---|---|---|---|---|
| | $\times 10^{32}$ erg | $\times 10^{32}$ erg | $\times 10^{32}$ erg | | | | | | | N |
| 13 00:00 | 1.10 | 0.04 | 1.15 | 1.04 | 0.14 | 0.02 | 1.5 | 0.06 | 0.11 | |
| 13 00:20 | 1.14 | 0.06 | 1.20 | 1.05 | 0.11 | 0.03 | 2.0 | 0.05 | 0.10 | |
| 13 00:40 | 1.14 | 0.05 | 1.20 | 1.05 | 0.12 | 0.02 | 2.0 | 0.05 | 0.09 | |
| 13 01:00 | 1.18 | 0.07 | 1.25 | 1.06 | 0.10 | 0.02 | 2.0 | 0.04 | 0.09 | |
| 14 23:40 | 4.97 | 0.45 | 5.43 | 1.09 | 0.06 | 0.02 | 5.0 | 0.10 | 0.13 | |
| 15 00:00 | 4.90 | 0.72 | 5.62 | 1.15 | 0.05 | 0.02 | 5.5 | 0.06 | 0.12 | |
| 15 01:20 | 4.86 | 0.58 | 5.44 | 1.12 | 0.07 | 0.05 | 4.5 | 0.04 | 0.11 | |
| 15 01:30 | 4.86 | 1.55 | 6.41 | 1.32 | 0.08 | 0.15 | 5.0 | 0.05 | 0.10 | |
| 15 01:40 | 4.94 | 1.07 | 6.01 | 1.22 | 0.12 | 0.20 | 4.5 | 0.04 | 0.11 | |
| 15 02:25 | 4.98 | 0.14 | 5.11 | 1.03 | 0.09 | 0.01 | 3.0 | 0.03 | 0.07 | |
| 15 02:45 | 4.97 | 0.25 | 5.23 | 1.05 | 0.08 | 0.02 | 3.0 | 0.03 | 0.09 | |
| 15 03:06 | 5.02 | 0.34 | 5.35 | 1.07 | 0.07 | 0.01 | 3.5 | 0.03 | 0.08 | |
| 15 04:00 | 5.11 | 0.42 | 5.53 | 1.08 | 0.07 | 0.02 | 3.0 | 0.07 | 0.10 | |
| 15 04:19 | 5.18 | 0.28 | 5.45 | 1.05 | 0.08 | 0.02 | 3.5 | 0.05 | 0.09 | |

## 4. Pre- and post-flare models

Having validated the QGR method for quiescent conditions, we now apply it to assess changes in the coronal field during a major flare. The chosen event is the "Valentine's day flare", the X2.2 class flare that occurred in the active region examined in Section 3, AR 11158, around 2:00 UTC on 15 February, 2011 (e.g., Tziotziou et al. 2013).

Active region AR 11158 exhibited repeated explosive activity and evolution in coronal loops around the time of the onset of the flare. The light curve and the running difference "light curve" for this time are shown in the right column in Figure 2. These light curves demonstrate that in the time interval between the first and the last time in the sequence of the QGR models there were three bursts of activity, which are illustrated in Figure 6.

In modeling the field, we focus on time intervals when no flaring activity is ongoing. Specifically, we choose two time intervals before the flare, one during the decay phase (after the loops stopped oscillating, according to Sun et al. 2012, but before the next eruption) and one after the flare. In order to provide an additional validation of the method, we construct several models within each time interval, typically 10-20 minutes apart. The expectation is that two models close in time should be similar to one another, to within the uncertainty of the method, provided there is no substantial activity between them. The instances to which we apply the QGR method are indicated by the arrows in the light curves in Figure 2. They span a five-hour interval centered on the X2.2 flare. This sampling in time allows us to characterize the general trend of the coronal magnetic field around the time of this flare.

We construct the models of the field maintaining the same field of view, spatial resolution and other parameters as in Section 3. The only exception is $\alpha_{thr}$, the "damping" parameter, which had to be modified for several models, as explained below. This parameter controls how much of a variation in $\alpha$ value along field lines is allowed, or, in another words, how much of departure from a force-free state is allowed in the solution. The default value is $\alpha_{thr} \approx 2 \times 10^{-6}$ arcsec$^{-1}$, the same as in Section 3. This value is chosen such that numerical errors of the estimated magnitude do not lead to changes in the fields between consecutive iterations. It can be increased for situations when input data are not consistent with a force-free solution, in attempts to find a solution as force-free as possible, that still matches the observations. This is needed at the three times immediately prior to the flare. For these three times we find that, as the scaling factor $f$ increases, the field becomes progressively more consistent with observations (i.e., $\overline{d}(f)$ decreases) up to the point when the solutions become oscillatory. The oscillations are much like those described in Malanushenko et al. (2012): the current-filling domain expands until it touches the boundary of the computational domain, where current on the field lines crossing the boundary is artificially set to zero, which shrinks the volume with currents, but the latter starts growing again as the volume constraints are reimposed. In attempts to find the minimum $\overline{d}(f)$ we therefore increase the value of $\alpha_{thr}$ for high-$f$ runs gradually until we find the smallest value of $\alpha_{thr}$ for which convergence is achieved. These values are $2 \times 10^{-4}$ arcsec$^{-1}$ for 1:20 UT and $1 \times 10^{-3}$ arcsec$^{-1}$ for 1:30 UT and 1:40 UT. In the seven other models, a minimum value of $\overline{d}(f)$ is readily found for the default value of $\alpha_{thr} = 2\times10^{-6}$ arcsec$^{-1}$.

Visual inspection, and quantitative testing, indicate that field lines of the QGR models at all times match the shapes of coronal loops inferred from EUV observations, with a reasonable degree of accuracy. Figure 7 labels distinctive bundles of loops in AR 11158, referred to as bundles A-G below. Figure 3 and the on-line only Figures S3 and S4 illustrate that the QGR models successfully reproduce these structures in most cases. Note, in particular, how the structures B and C are mostly absent in the images for the the decay phase and postflare; this is reproduced in QGR modeling. The only exceptions when the individual structures are not well reproduced are structure C in the first pre-flare series, structure C in the immediate pre-flare series and the outer-most region of the structure D in the pre-flare series and in the first model of the immediate post-flare series.

We also find that the major magnetic structures in the QGR models have consistent signs and values of twist in the QGR modeling at different times. This is illustrated in Figure 3, as well as in the on-line only Figures S3 and S4. In particular:

- the field in the double-arched loop structure A is consistently found to have a large and negative twist;
- the central sigmoid bundle B consistently has a large and positive twist in all pre-flare models, while in the decay phase and in the post-flare models, the loops at that location have a weaker, but still positive twist;
- the reverse-sigmoid bundle C consistently has a negative twist; in the decay and post-flare phases the structure is mostly absent;
- the peripheral bundles D and E are consistently found to have a small positive twist;
- the bundles F and G, which connect the active region with the surrounding corona, have a negative twist, and mixed signs of twist, respectively, and the individual loops in the bundle G are generally weakly twisted.

We also find that the structure of currents around the time of the flare is reproduced with remarkable consistency. This is illustrated in Figure 4, as well as in the on-line only Figures S5 and S6. These figures show the vertically integrated current density magnitude for the domains where currents are parallel (left panels) and anti-parallel (right panels) to the field. A gradual build-up of currents is observed in both domains prior to the time 1:30 UT, and a rapid decay is observed from 1:40 UT to 2:40 UT, with smaller changes from then on.

We present also a quantitative assessment of the consistency of the model results with time, applying metrics from Section 3 and following a similar approach. (See the right column in Figure 5 for the numerical results.) The findings are that:

- according to all metrics, the field models *after* 2:00 UT resemble each other more than they do the corresponding potential fields and the models which are closer in time are the most similar to each other. This behavior is similar to the results for the quiescent phase;

- there are substantial differences between the field models in the immediate pre-flare phase, *before* 2:00 UT, even between the field models close in time. Prior to 2:00 UT models that are closer in time are less similar to each other than models that are further apart;

- the largest changes in the field are observed between 1:30 UT and 2:40 UT, which is consistent with the timing of the X2.2 flare. The field models then are, according to some metrics, more dissimilar to each other than they are to the corresponding potential fields.

To summarize, the QGR models of AR 11158 are remarkably consistent in the quiescent and the post-flare times, but immediately prior to the flare the models are less consistent. It is noteworthy that the $C_{CS}$ and $C_{vec}$ metrics, which are sensitive to the difference in direction, demonstrate better results than $1 - E_{m1}$ and $1 - E_{n1}$, which are also sensitive to the difference in magnitude between the two fields. We take this to mean that the QGR models are more consistent in determining the direction of the field than the magnitude of its non-potential component.

Table 1 summarizes some of the properties of the model fields from our analysis. In particular, note the rapid increase in free energy between 1:20 UT and 1:30 UT, and a rapid decrease by 2:25 UT. The potential energy $E_P$ changes relatively slowly by comparison with the total energy $E$. Figure 8 plots the variation in the energies as a function of time. Also note the elevated values of the force-freeness parameter $\xi$ in the last three models immediately prior to the flare, which are due to the increased value of $\alpha_{thr}$ used to compute these three models, as we explain above (also shown in Figure 8). Table 1 also demonstrates that for all of the examined times, QGR models are a better match to the observed loops than potential field models. It also appears that prior to the flare, and right before the next eruptive event, our procedure of fitting loops has a systematically lower success rate (51%-66%, compared to about 75% success rate in the quiescent time, and the post-flare series).

We further investigate the discrepancies between field models immediately prior to the flare. For each time, we analyze individual solutions with different $f$ values. The left column in Figure 9 shows how $\overline{d}$, $\xi$ and $E_F$ change with $f$. The plots of $\overline{d}$ and $\xi$ (upper and middle panels) are shown as a function of free energy $E_F = E - E_P$. The lower panel shows that free energy increases with $f$ in each case. For comparison, the right column in Figure 9 shows the same set of plots, for the models constructed immediately after the flare. In every case (except at 1:20 UT) the free energy increases with $f$ up to a certain critical $f$ value. Once this value is reached, the QGR solution is found to be unstable unless $\alpha_{thr}$ is increased, which increases the value of $\xi$. Below this critical value, $\overline{d}(f)$ decreases monotonically, and than increases with increasing $f$. The model behaves differently at 1:20 UT when $f > 4$. The solution in that case is found to be quite unstable. We believe that this might be a boundary artifact, for the following reasons. At larger $f$ values the bundle D becomes tilted rightwards and touches the side boundary at 1:20 UT. As no currents are allowed to leave the domain through the side boundaries, the current in the bundle D decreases, which makes the bundle less tilted, but as the constraints in the volume are reimposed, the tilt increases again. We notice that in the 1:30 UT and 1:40 UT solutions, the bundle D is sheared more, to the extent that it does not touch the northern side boundary again. We therefore believe that in a larger computational domain, the solution at 1:20 UT would converge with a smaller value of $\alpha_{thr}$ and therefore with a smaller $\xi$ value to a solution with larger free energy, like it does at the two later times.

The minimum of $\overline{d}(f)$ appears rather shallow in Figure 9 for most times; while we report the uncertainties in free energy based on this figure, we believe the actual uncertainties are likely to be smaller. We examine the solutions for different $f$ values, as shown in the on-line only Figure S7 and notice that despite small difference in $\overline{d}(f)$, the field lines of the solution have quite different shape and the shear angle of the bundles D and E changes rather noticeably with $f$. The metric $\overline{d}$ was initially designed to quantitatively describe such behavior. It is possible that in future studies, $\overline{d}$ will require revision in formulation. Meanwhile, we base our conclusions on the existing metric, but note that the actual uncertainties in determining $f_{best}$ and therefore the uncertainties in the free energy may be smaller than we report.

Based on Figures 8 and 9 and Table 1 we estimate that during the X-class flare, the free energy of the field has decreased by $\Delta E_F \approx (1 \pm 0.5 \times 10^{32})$ erg. This estimate is made assuming that the QGR models just before the flare (or just after the flare), are different realizations of roughly the same state of the corona before (or after) the flare. We note, however, that the free energy $E_F$ monotonically increases from 2:25 UT onwards, along with the potential field energy $E_P$. We take it to be a consequence of the fact that ropes in the central part of the AR 11158 continue to emerge. This is supported by existing studies (e.g., Tziotziou et al. 2013). In this case, the estimated energy change is based on values $E_F \approx 1.3 \times 10^{32}$ erg just before the flare (an average for the 1:30 UT and 1:40 UT models, discarding the 1:20 UT model for the reasons explained above) to $E_F \approx 0.1 \times 10^{32}$ erg just after the flare (2:25 UT model).

We also test the uniqueness of the QGR solutions at a given time by performing the following "permutation test". The solution at the earlier time is used as the initial guess for the later time, and *vice versa*. The QGR iteration also requires an initial guess for $\alpha$ in the volume. In the permutation test, the initial guess for $\alpha$ in the volume is simply the $\alpha$ cube from the other QGR solution. When the iterations are initialized with the potential field, the initial guess for $\alpha$ in the volume is determined by $\alpha$ at the nearest loop trajectory.

The results of the permutation test are summarized in Table 2. We conclude that QGR for the pre-flare states is sensitive to the initial guess, but the free energy estimates are within the range reported, $E_F \approx (1.1 \pm 0.5) \times 10^{32}$ erg. Table 2 indicates also that the quality of the match to the coronal loops is slightly worse when the initial guesses are not a potential field with a piecewise-constant $\alpha$ field determined by the constraints in the volume. The parameter $\xi$, which measures how force free the field is, is lower when the initial guess is taken from the other QGR solution.

This may be caused by the solutions being closer to having a critical value of $f$, as explained above. In this situation a small increase in $f$ results in a large increase of $\xi$.

Table 2: Results of the permutation test, where the solution for one time is used as the initial guess for modeling at another time.

| Constraints in the volume and the lower boundary | Initial guess for $\mathbf{B}$ | $E_F$ | $E_{tot}$ | $E/E_P$ | $\xi$ | $f_{best}$ | $\overline{d}_{best}$ |
|---|---|---|---|---|---|---|---|
| | | $\times 10^{32}$ erg | | | | | |
| 2011-02-15 1:20 | Potential field at 1:20 | 0.58 | 5.44 | 1.12 | 0.05 | 4.5 | 0.04 |
| 2011-02-15 1:20 | QGR solution at 1:30 | 1.23 | 6.09 | 1.25 | 0.04 | 5.5 | 0.05 |
| 2011-02-15 1:30 | Potential field at 1:30 | 1.55 | 6.41 | 1.32 | 0.15 | 5.0 | 0.05 |
| 2011-02-15 1:30 | QGR solution at 1:20 | 0.85 | 5.71 | 1.17 | 0.02 | 3.75 | 0.06 |

Finally, we examine the structure of the field for the QGR models of the field just before, and just after the flare. Figure 10 shows the changes in the core of the active region (corresponding to the loop bundle B in Figure 7) between 1:30 UT, when the energy in the modeled field is a maximum, and 2:45 UT, which is the first modeling time after the flare. AIA 304Å images are shown for comparison.

At 1:30 UT, there is a sigmoid flux bundle along the polarity inversion line. We take a vertical slice across the polarity inversion line and trace field lines from each point on this slice, integrating current magnitude along field lines. We find that the currents are localized in a relatively small region of size about 6Mm $\times$ 4Mm in cross section in the vertical plane. The field lines in this region are twisted by $Tw \approx 0.6$ full turns, estimated from the solution using the twist number defined, for individual field lines, as $Tw = \alpha L/4\pi$, where $L$ is a field line's length and $\alpha$ is the value on this field line (Malanushenko et al. 2009a). For thin flux tubes, $Tw$ is the total number of turns a field line makes about the axis of the tube. The core of the high-current region is found to be twisted by $Tw \approx 0.9$ turns, which is consistent with the result of Sun et al. (2012), who found a twist $Tw \approx 0.6$ on average in the bundle and up to $Tw \approx 1$ in the center. Sun et al. (2012), however, found that the peak of the twist at approximately the same location along the polarity inversion line is reached at a height of $\approx 2$ Mm, while in our model the peak occurs at height $\approx$ 4Mm.

In the QGR models constructed for 2:45 UT, the entire localized current structure is gone. The currents are weaker and are more uniformly distributed. The same applies to the values of twist. It is noteworthy that the structure which in our model disappears after the flare is similar in shape and location to the filament which was observed to erupt (e.g., Schrijver et al. 2011).

Such a value of $Tw$ might be indicative of that the twisted structure in the core of the region is close to becoming kink unstable. The critical value of $Tw$ for a kink instability onset has been estimated to be $Tw \approx 1.4 - 1.7$ depending on a particular configuration (Hood & Priest 1979; Malanushenko et al. 2009a). It is plausible that the QGR modeling underestimates the actual amount of twist: it is derived from a Grad-Rubin iteration, which is proven to converge only for relatively weakly twisted fields (more specifically, for $\alpha$ sufficiently small, see Bineau 1972). This might explain the difficulties with convergence of the pre-flare models.

## 5. Conclusions

This paper demonstrates the utility of the QGR method when applying it to a time series of line-of-sight magnetogram and coronal data associated with AR 11158. In particular, we verify that the method yields self-consistent results when modeling the field in the quiescent phase of this growing active region on February 13, 2011. We demonstrate that the QGR models of the field: (1) are force-free according to our metrics; (2) reproduce the shape of observed coronal structures better than potential field models; (3) are closer to each other than they are to the corresponding potential field models using quantitative measures; and (4) change slowly with time, consistent with the observed slow evolution of the photospheric field and the absence of eruptive activity in this period.

We also apply the QGR algorithm to the field before and after a major flare. The test case chosen is active region AR 11158 around the time of the X2.2 "Valentine's Day Flare" of February, 15th, 2011. For each time selected for modeling, we construct another model within a short time interval of the order of a typical lifetime of coronal loops, to provide additional validation to our measurements. We find that close to the time of a flare, the models become substantially more sensitive to small variations in the input data, but that models close in time become consistent to each other again several hours after the event. We also find that the spatial distribution of electric currents is consistently reproduced by models close in time both before and after the event. We show that the uncertainty in the energy estimates is small enough to observe a significant drop in the free energy during the studied event.

The energy estimates of the QGR models shown here indicate a rapid decrease of free energy by about $\Delta E_F \approx (1 \pm 0.5) \times 10^{32}$ ergs during the flare, which is consistent with the expectation of the energy release in an X2 flare, as summarized by Schrijver et al. (2012). They reviewed estimates of total energies involved in solar and stellar explosive events and concluded that an X4 flare would require an energy budget of $\sim 5 \times 10^{32}$ ergs, subject to an uncertainty of a factor of about three.

Our energy estimates are affected by the saturation in the HMI line-of-sight magnetograms. The latter underestimate the field strength for strong fields, i.e., where $|\mathbf{B}| > 1$ kG (Hoeksema et al., 2013). The spectropolarimetric inversions of the HMI data give field strengths larger by at least a factor of 1.2-1.3. Magnetic energy is a volume integral of $B^2$, and therefore a field stronger by roughly a constant factor of 1.2-1.3 will have energy measurements larger by a factor of 1.4-1.7. Given that the fields over 1kG are expected to contribute most to the large scale coronal field, we believe that by not correcting for the saturation factor, we underestimate energy content in the corona. This effect causes our estimate of potential energy, $E_P \approx 5 \times 10^{32}$ ergs around the time of the flare, to be consistent with that reported by other studies that use line-of-sight data derived using an MDI-like algorithm (Tarr et al. 2013), while studies that use vector field data from spectropolarimetric inversions estimate the potential field energy to larger by about a factor of 1.6 (e.g., Sun et al. 2012, reported $E_P \approx 8 \times 10^{32}$ ergs). It is interesting that despite the influence of the line-of-sight saturation bias, our modeling suggests an energy drop larger than that in Sun et al. (2012). They estimated $\Delta E_F \approx 0.3 \times 10^{32}$ ergs, which, in our rough scaling estimate, would be even smaller should one rescale the vector field to match the line of sight field (or that is to say, if we used the lower boundary data of the same scaling for stronger fields). Tarr et al. (2013) estimated $\Delta E_F \approx 1.7 \times 10^{32}$ ergs, which is larger than our estimate. We believe that our results are not inconsistent with those of Tarr et al. (2013), as their estimate of the drop in the free energy is obtained by subtracting two numbers: the pre- and post-flare *lower bounds* to the free energy, obtained via the minimum current corona analysis (Longcope 2001).

We also find that in the pre-flare QGR field models, there is a strongly twisted sigmoid bundle of flux along the polarity inversion line in the active region core. The location and shape of that structure are consistent with the observed location and shape of the filament which erupted at the time of the flare (Schrijver et al. 2011). We find that the strongly twisted flux bundle is absent in the post-flare models, which is again consistent with the observations.

Overall, we conclude that the QGR method successfully models the solar coronal field over active regions. Based on this study the method is suitable for evaluating parameters of the field critical for space weather forecasting, including the free energy available for release, the spatial distribution of electric currents, and the features of the magnetic field such as twisted structure in the core of the pre-flare AR 11158.

This work was supported by NASA contract NNG04EA00C for the SDO Atmospheric Imaging Assembly. The work benefited from discussions by the International Space Science Institute (ISSI) in Bern during meetings of International Team 238, “Nonlinear Force-Free Modeling of the Solar Corona: Towards a New Generation of Methods”. We gratefully acknowledge the support provided by ISSI.

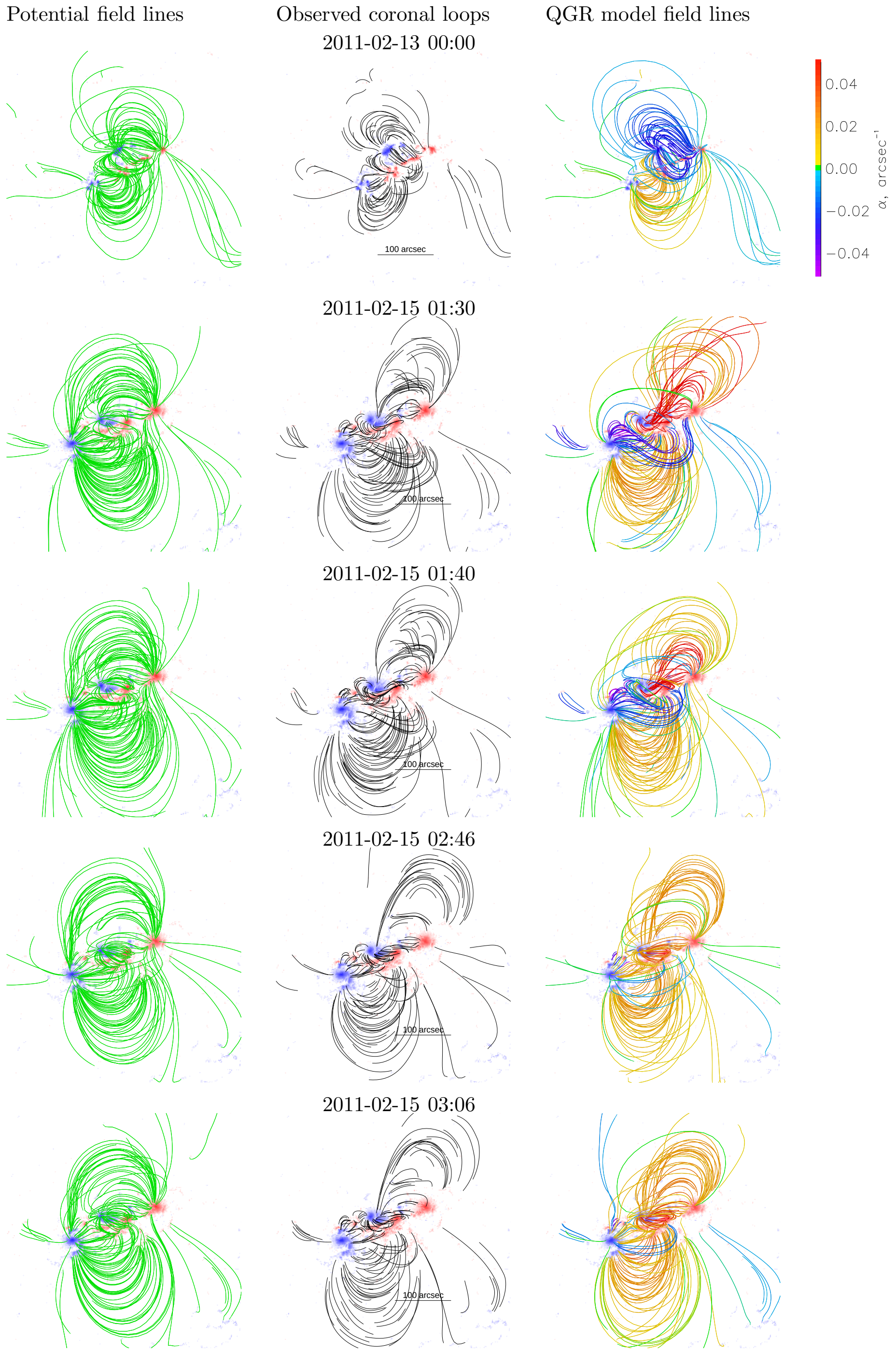

Fig. 3.— Coronal loops, traced from coronal images (middle column), best-fitting field lines of a potential field (left column) and GR solutions (right column) over the magnetograms (the background red/blue images). The fitting procedure is described in Section 2. The colors correspond to the value of $\alpha$.

Vertically integrated density of currents in the two sub-volumes

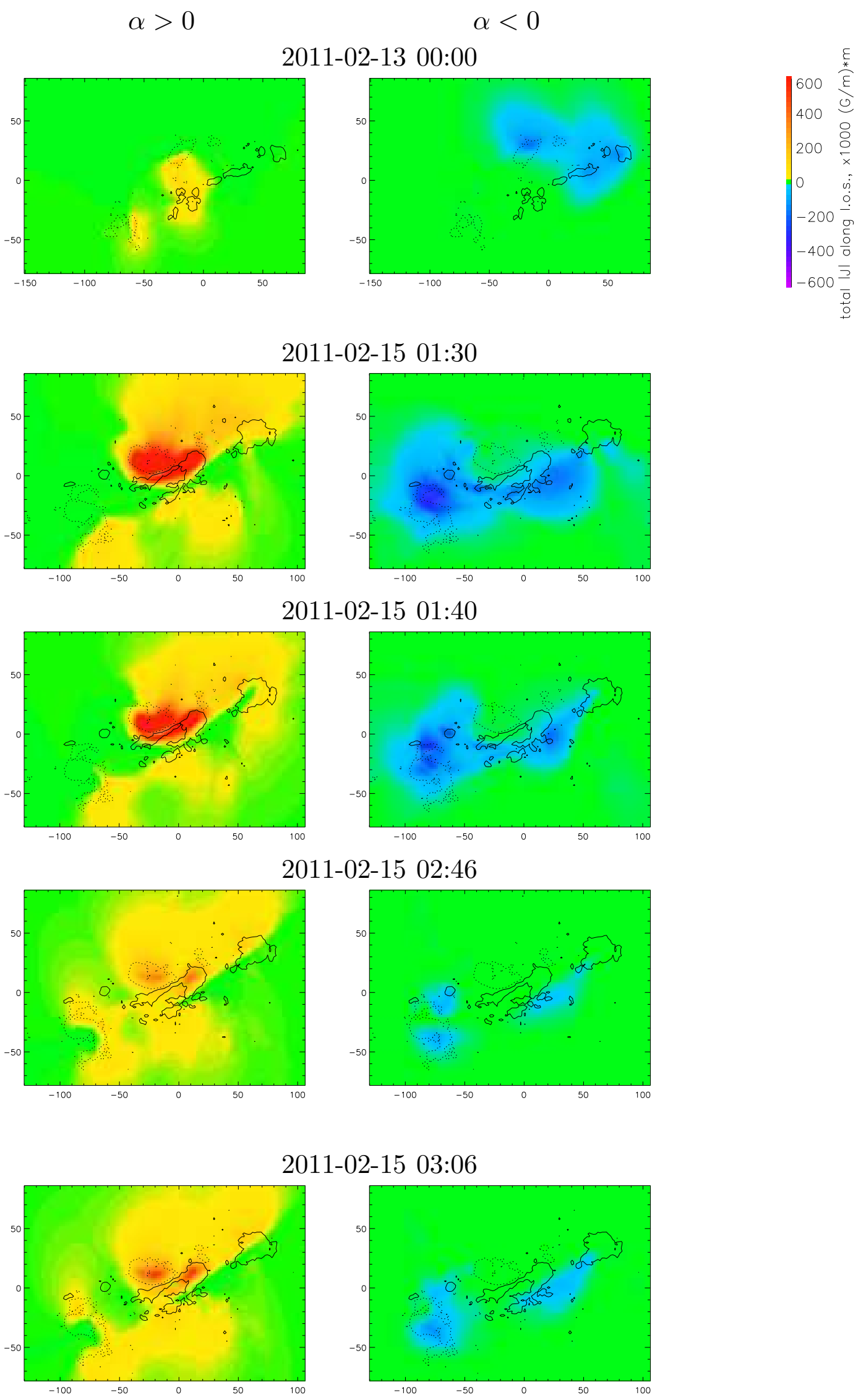

Fig. 4.— The vertically integrated current density in the central portion of our domain, for currents aligned parallel (left column) and antiparallel (right column) to the field.

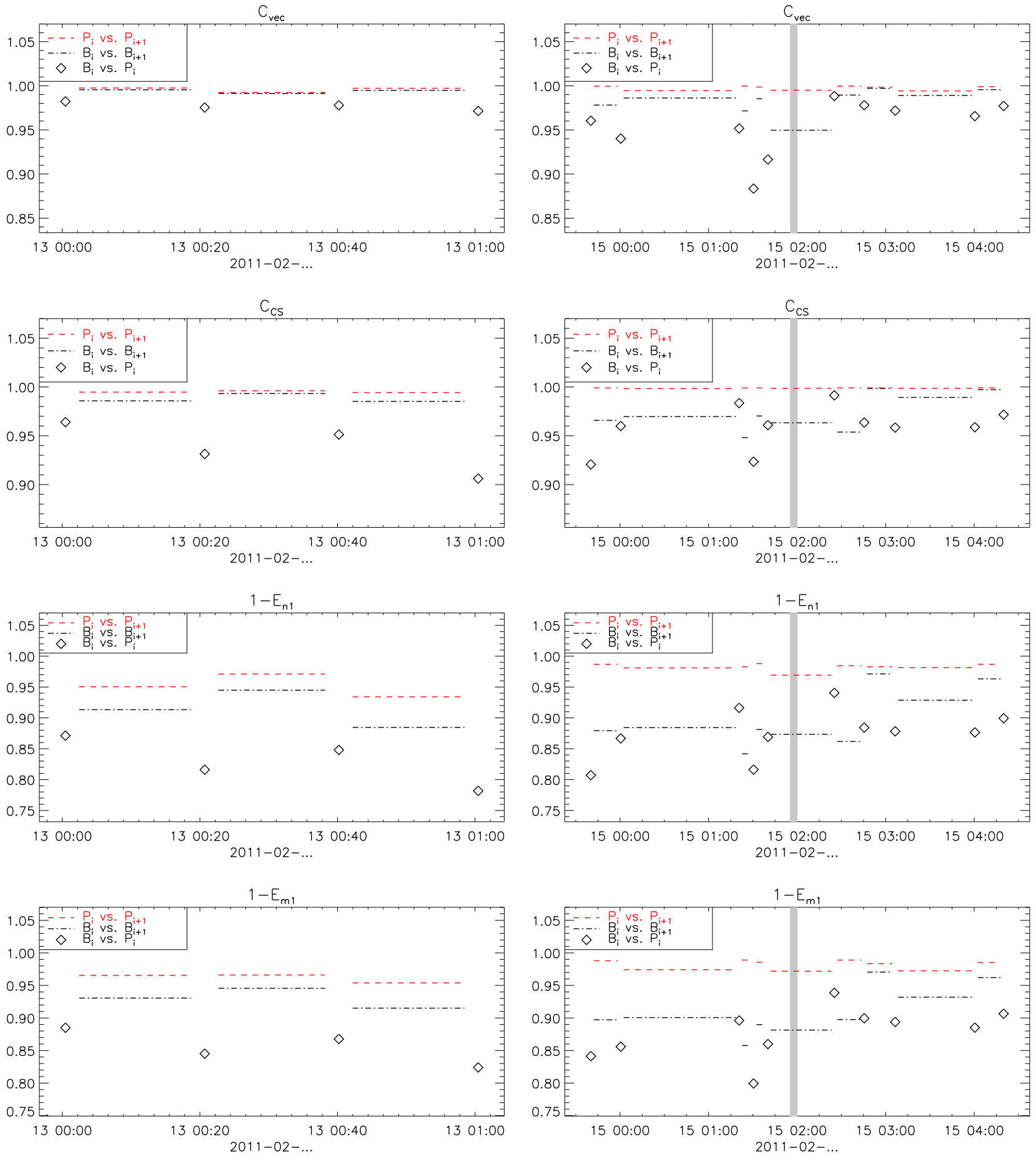

Fig. 5.— Measures of the similarity between pairs of consecutive in time QGR models (black dashed lines), the corresponding potential fields (red dash-dotted lines) and between a QGR model and the corresponding potential field (diamonds) as measured by the metrics $\mathrm{C_{CS}}$, $\mathrm{C_{vec}}$, $1 - E_{m1}$ and $1-E_{n1}$. Left column: quiescent time. Right column: time around the flare (gray band indicates the peak time of the flare). Note the following: (1) the metrics for the consecutive potential fields are of order 0.95 or higher; (2) in the quiescent and the post-flare phases, the metrics resulting from the comparison amongst the set of QGR models are almost as high as those from the potential field set; (3) in the quiescent and the post-flare phases, when comparing the QGR models to the potential field models at each time, the metrics are all lower than the comparisons between potential or QGR models to themselves.

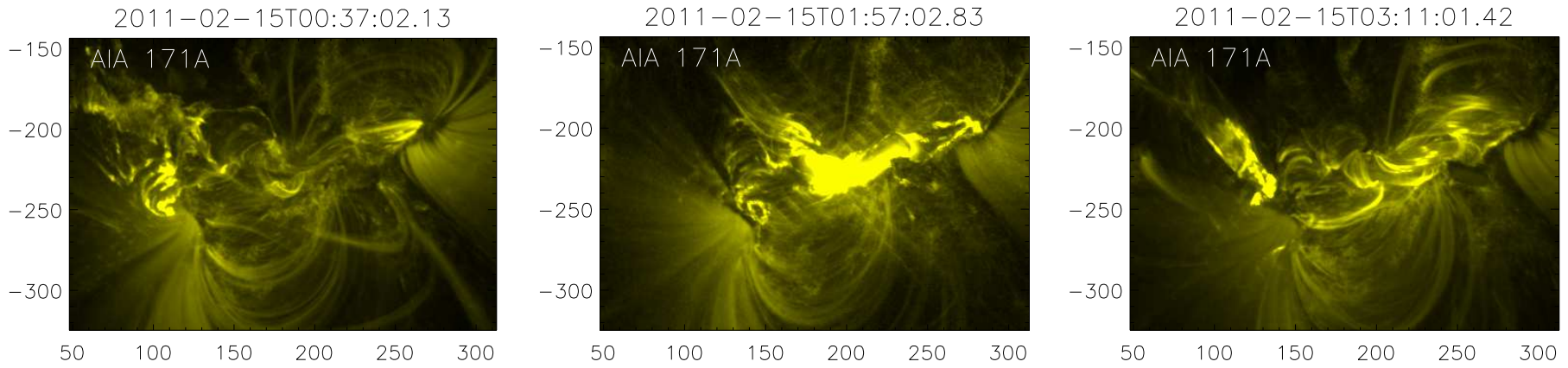

2011-02-15 00:27–00:50
Double-arch eruption in the North-East of the active region.

2011-02-15 01:45–03:00
X2.2 class flare in the center of the active region. Noticeable oscillations in the large-scale loops in the north of the region end at around 02:24, and the saturation in the center of the region continue until around 03:00.

2011-02-15 03:07–03:43
Spray surge in the North-East of the active region, followed by slow descent of cool dense material.

Fig. 6.— AIA 171Å images of the eruptive events during the examined time frame, together with their approximate duration and description.

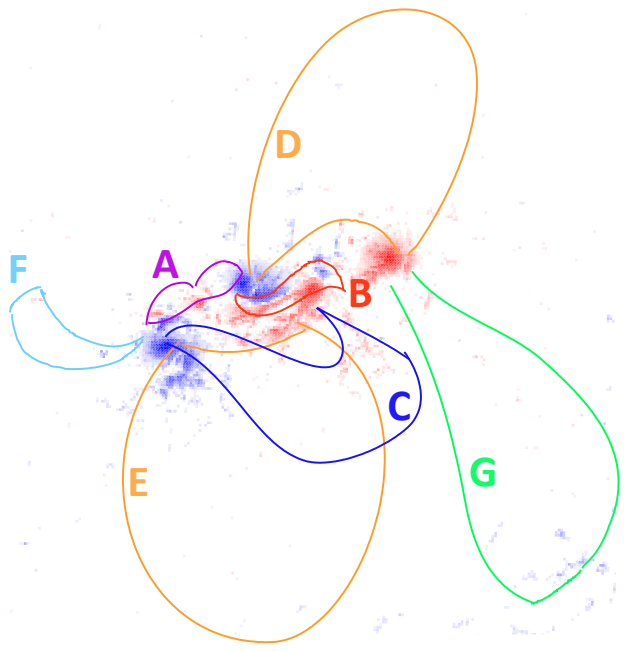

Fig. 7.— A sketch of prominent loop structures visible in AIA images of AR 11158 around Feb 14-15. We use the letter notation shown to refer to these structures.

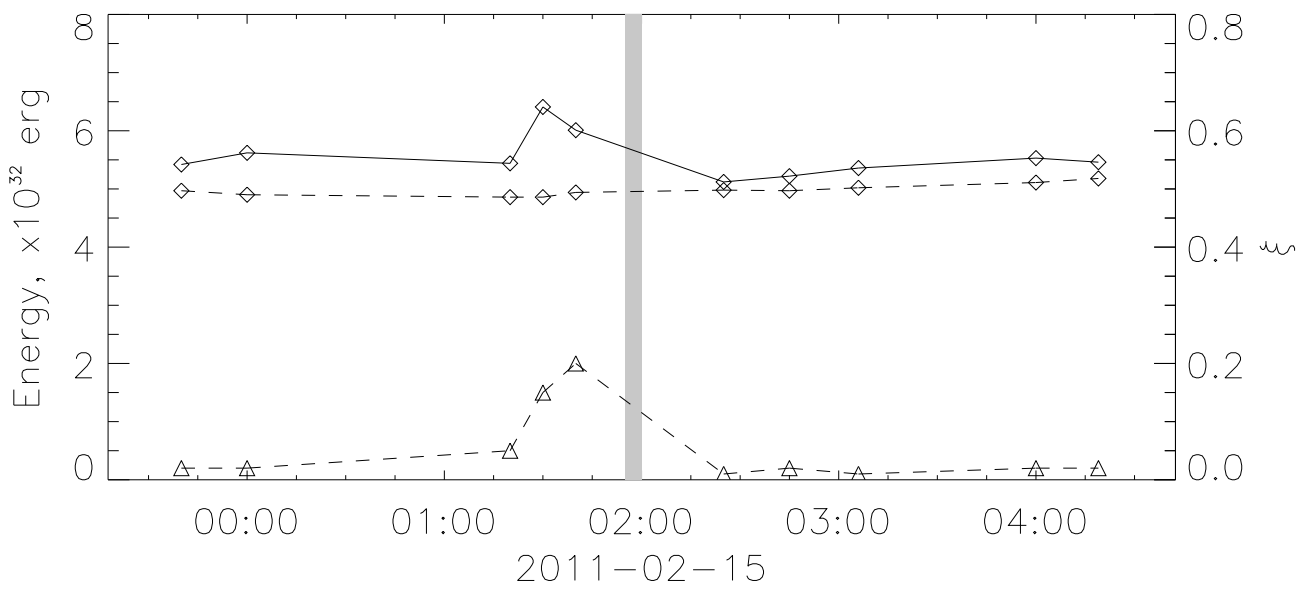

Fig. 8.— The energies for the field models for AR 11158 around the time of the X2.2 flare (gray line). The total energy $E$ (solid line with diamonds) and the potential field energy $E_P$ (dashed line with diamonds) are shown together with the force-freeness parameter $\xi$ (dashed line with triangles).

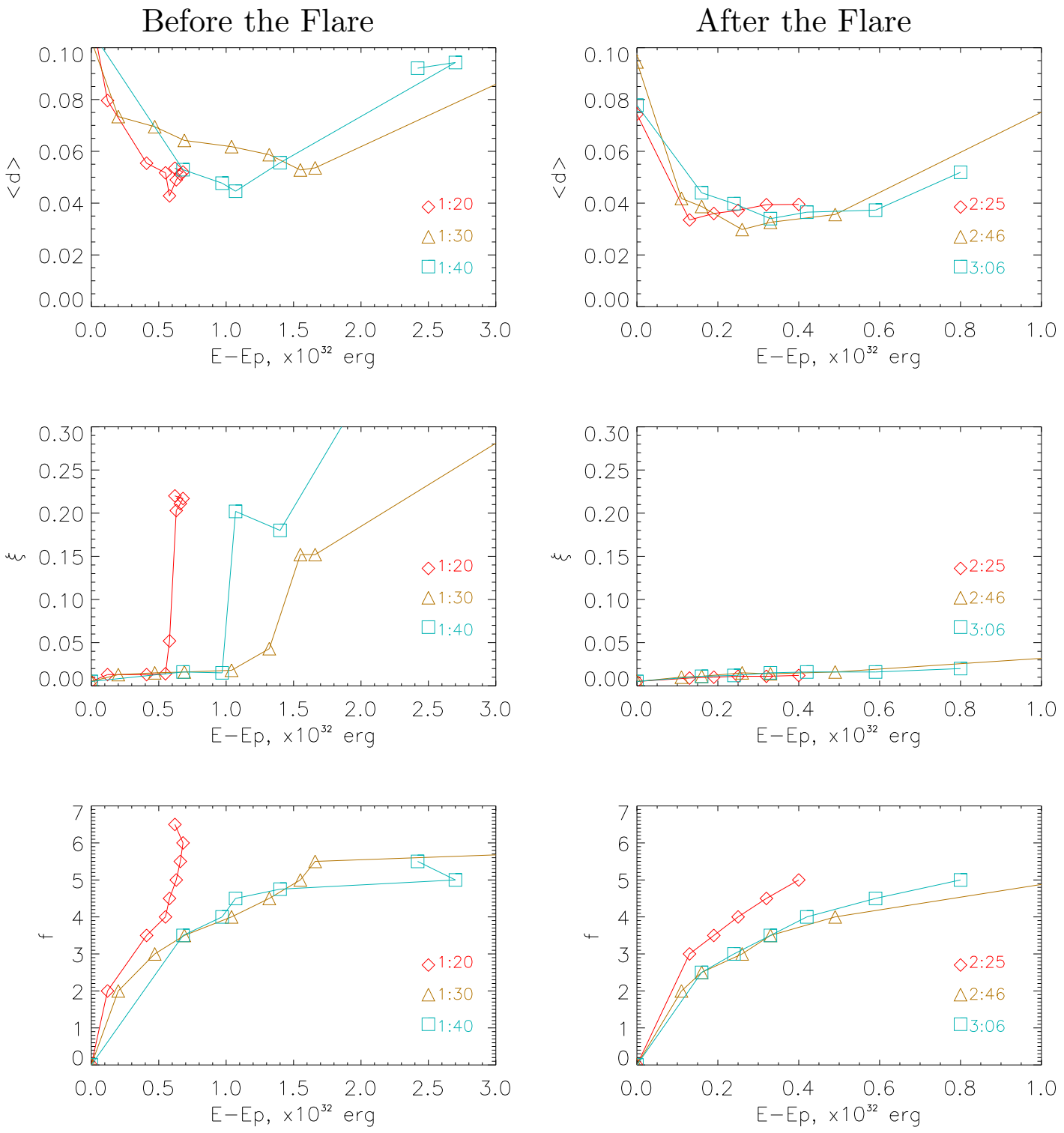

Fig. 9.— The quality of the fit to coronal loops (top panels) and the metric $\xi$ measuring how force free the solution is (middle panels) for QGR models constructed with different $f$ values. The free energy, $E_F = E - E_P$, depends on $f$ as shown in the lower panels. The left column shows the three QGR models right before the flare and the right column shows the three QGR models right after the flare.

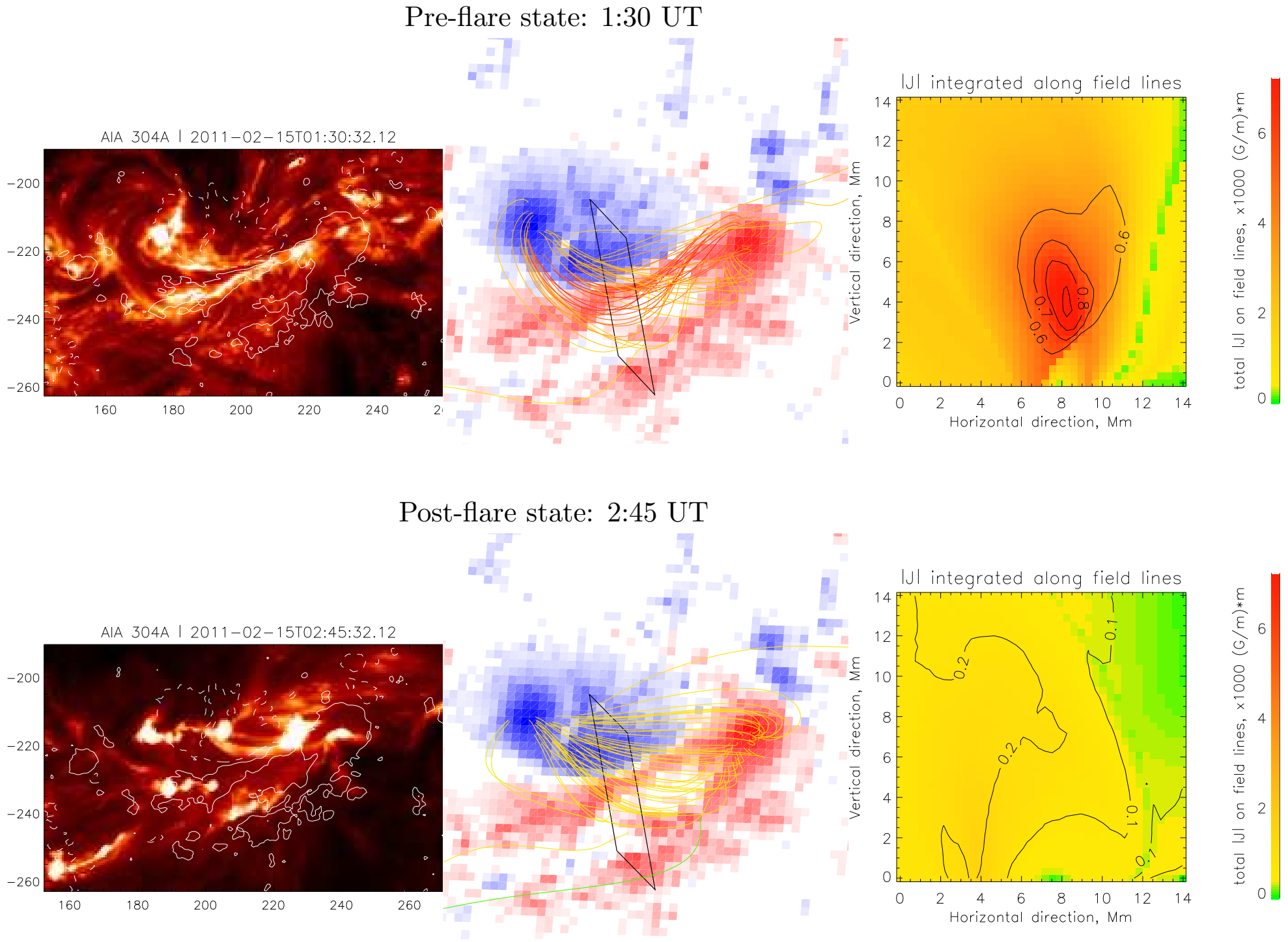

Fig. 10.— A close-up of the center of the active region at the pre- and post-flare times, showing drastic change in the field structure along the polarity inversion line. This region is referred to as structure B in Figure 7. Left column: AIA 304Å images for reference. Middle column: close-ups of field lines in the QGR models viewed along the line of sight. A rectangular vertical slice is shown and selected field lines initiated from this slice are shown, colored by total current along them. Right column: electric current magnitude, integrated along field lines, for a vertical central slice of the QGR models, indicated by the vertical rectangle drawn in the middle column panels. The contours show values of the measure of twist, $Tw = \alpha L/4\pi$. It is evident that there is a highly twisted structure with $Tw \approx 0.9$ total turns in the pre-flare model, which is gone after the flare.